\documentstyle[epsf]{llncs}
\title{Some Techniques for the Measurement of Complexity in Tierra}
\author{Russell K. Standish}
\institute{High Performance Computing Support Unit\\
University of New South Wales\\Sydney, 2052\\Australia\\
R.Standish@unsw.edu.au\\http://parallel.hpc.unsw.edu.au}

\begin{document}
\maketitle

\begin{abstract}
Recently, Adami and coworkers have been able to measure the
information content of digital organisms living in their {\em Avida}
artificial life system. They show that over time, the organisms behave
like Maxwell's demon, accreting information (or complexity) as they
evolve. In {\em Avida} the organisms don't interact with each other,
merely reproduce at a particular rate (their fitness), and attempt to
evaluate an externally given arithmetic function in order win bonus
fitness points. Measuring the information content of a digital
organism is essentially a process of counting the number of genotypes
that give rise to the same phenotype.

Whilst Avidan organisms have a particularly simple phenotype, Tierran
organisms interact with each other, giving rise to an ecology of
phenotypes. In this paper, I discuss techniques for comparing pairs of
Tierran organisms to determine if they are phenotypically
equivalent. I then discuss a method for computing an estimate of the
number of phenotypically equivalent genotypes that is more accurate
than the ``hot site'' estimate used by Adami's group. Finally, I
report on an experimental analysis of a Tierra run.
\end{abstract}

\section{Introduction}

The issue of what happens to complexity in an evolving system is of
great interest. In natural (biological) evolution, the naive view is
that life started simple, and evolved ever more complex life forms
over time, leading to that pinnacle of complexity, {\em homo
sapiens}. The end points of that process are of course fixed. In the
beginning, life must be simple. In our present era, there must exist
intelligent organisms (namely us) pondering over the mystery of how we
came to be. So the {\em anthropic principle} fixes the present day as
having complex lifeforms. There is nothing within the {\em Modern
Synthesis} of Darwinism that implies a steady interpolation between
these two end points. In fact it is even plausible that more complex
organisms than us existed in the past, but have since vanished into
obscurity. However, examinations of the fossil record over the
Phanerozoic (the last 550 million years of the Earth's history)
indicate almost no growth in complexity by a number of different
measures over that period, apart from an initial large jump at the
Cambrian explosion.\cite{McShea96}

The interesting thing is to ask what one might see if looking at
another evolutionary system apart from the one in which we
evolved. Would we see any growth in complexity at all? Since we don't
have an extra terrestrial biology to observe (a few Martian meteorites
aside), the only other systems available are Artificial Life systems
evolving within a digital computer such as Tierra or Avida. The Avida
group has reported measuring the information content (complexity) of
individual avidan organisms\cite{Adami98}, or rather a lower bound of
the organism's complexity. Their results are that this lower bound
increases over time for the maximally fit organism, thus showing
information accumulating as time progresses. One important critique of
this work, however, is that organisms do not interact directly with
each other, and in order to prevent evolution stagnating, an
externally imposed task (eg computing a logical operation) is added to
the system. Organisms are given ``fitness points'' depending on how
well they perform this task. This heavily weights the system in favour
for accruing information.

By contrast, in the Tierra system, the organisms interact with each
other, providing a rich array of possible (intrinsic) tasks for the
organisms to exploit. Since this is an evolving ecology with no
externally imposed task, the above critique does not
apply. However, the downside is that determining whether two genotypes
are phenotypically equivalent is considerably more complex. In some
work a couple of years ago\cite{Standish97b}, I studied the phenotypic
properties of Tierran organisms to build up a picture of the genotype
to phenotype landscape. A Tierran organism's phenotype can be 
characterised by a couple of numbers for each possible pairwise
interaction in the ecology. Multiway interactions are ignored in this
study, as experience has shown them to be relatively rare.

\section{Complexity of a Digital Organism}

The information content of a string is given by the difference between
the maximal Shannon entropy of that string (i.e. considering the string
to be random, or devoid of information), and the entropy given by
assuming that the string codes for some phenotype $p$:\cite{Adami98,Layzer88}
\begin{equation}
I(g) = H(g) - H(g|p) = \ell - \log_{32}N
\end{equation}
where $\ell$ is the length of the genotype (in instructions), and $N$
is the number of genotypes that give rise to the same phenotype
$p$. The base, 32, refers to the number of instructions in the Tierra
instruction set. If $N\approx32^\ell$ (ie a completely random
sequence), then $I(g)=0$. Similarly, if $N=1$ (there is only one
genetic sequence encoding a genotype, or no redundancy), then $I(g)=\ell$.

The most obvious way to compute $N$ is to search all $32^\ell$
genotypes for equivalent phenotypes. However, this is an enormous
number of strings to check, and computationally infeasible. Adami
recognised this problem, and took the approach of counting the number
of volatile sites $v$ (sites that vary amongst phenotypic equivalents),
and approximating $N\approx 32^v$. In one sense this is an
overestimate of $N$, so they argue that this gives a lower bound to
the information $I(g)$. In another sense, however, it is not strictly
a lower bound. If it turns out that fixing one of the volatile sites
to a particular value allows one of the fixed sites to vary without
altering the phenotype, then this would be not be counted in the
$N$. so what we have is really an overestimate of an underestimate.

The same criticism applies to this work. We can estimate the above
mentioned estimate fairly accurately, more precisely we can find the
size of the neutral network\cite{Kauffman95,Reidys-etal96, Schuster97}
connected by one-site neutral mutations to $g$. However, the
possibility remains that there are other neutral networks of $g$ that
aren't connected by single site mutations to $g$. Probably the most
efficient way of finding these is by using a genetic algorithm to
explore genotype space, i.e. run Tierra for a long time to see what it
discovers! The way we use this in our experiment is to keep a list of
neutrally equivalent organisms that Tierra discovers. As we explore
the neutral network connected to $g$, we eliminate items from the list
that we come across. The remaining names on the list can then be used
as seeds to start the process again.

In this work, we use two different techniques to measure $N$. The
first is a Monte Carlo random sampling technique to estimate the
proportion of the $32^v$ strings found by varying the volatile sites.
The second technique, which we use in conjunction with the Monte Carlo
approach mentioned above, is to walk the neutral net. The Monte Carlo
technique works well when the density of neutral variants is fairly
high, whereas the latter technique is best on sparse networks. A decision
on which technique to use for which site is based on estimated
densities of neutral variants.

\section{Establishing Phenotypic Equivalence}

Equation (4) of \cite{Standish97b} presents the dynamical equations of
two species of Tierran organisms interacting. The precise form of the
dynamics is not important here, however the phenotype of the organism
can be characterised by its interactions will all other possible Tierran
phenotypes. Since it is impossible to have the complete set of all
possible Tierran organisms, those organisms generated during a run of
Tierra are used. Since Tierran organisms coevolve, the most important
organisms should be contemporaneous with the test organism. The
following characteristics are saved for each pair of organisms:
\begin{enumerate}
\item The outcome of the tournament. This may be one of the following:
\begin{description}
\item[infertile] The test organism never calls the divide instruction,
or does not produce any recognisable progeny (essentially still born)
\item[once] The organism produces progeny once, but then never repeats
the act.
\item[repeat] The organism continuous reproduces the same progeny. For
this purpose we ignore what is produced first time around, as this
will be swamped by number latter progeny.
\item[nonrepeat] The organism continuously reproduces, but the progeny
is either different each time, or the CPU is in a different state each
time the divide instruction is called - thus can't be guaranteed to
reproduce ad infinitum.
\end{description}

\item The name of the progeny organism. This is usually identical to
the parent, but may another type in the case of symbiosis or
parasitism.

\item The number of timesteps it takes to reach the first divide
instruction ($\sigma_{ij}$), and the time it takes between successive
divide steps after that ($\tau_{ij}$).

\item The number of template matching operations made to the opposing
organism prior to the first divide ($\mu_{ij}$) and between successive
divides ($\nu_{ij}$).
\end{enumerate}

Two organisms are neutrally equivalent if they have identical
characteristics against all Tierran organisms. Once all organisms are
paired with each other, we can produce a list of phenotypically unique
organisms, which provides a smaller test list to pit trial mutants
against. We may also eliminate some noninteractive pairings prior to
simulation by trying to see if potential template matches could happen
between organisms. This still produces a fairly large list of test
organisms, so it is still computationally expensive. The high degree
of parallelism in this problem allows it to be attacked in reasonable
time on a parallel supercomputer.

A further refinement may be possible by producing an archetypal list,
perhaps by ignoring the ($\mu,\nu,\tau$ and $\sigma$) parameters. The
idea being that the archetypes contain a representative organism from
each niche of the ecology, and ignoring minor differences such as
reproductive rate. This would coarsen the approximation a little, but
will probably give an acceptable result. At present this idea has not
been tested.

\section{Interim Results}

\begin{figure}
\begin{center}
\mbox{}\epsfxsize=.7\textwidth\epsfbox{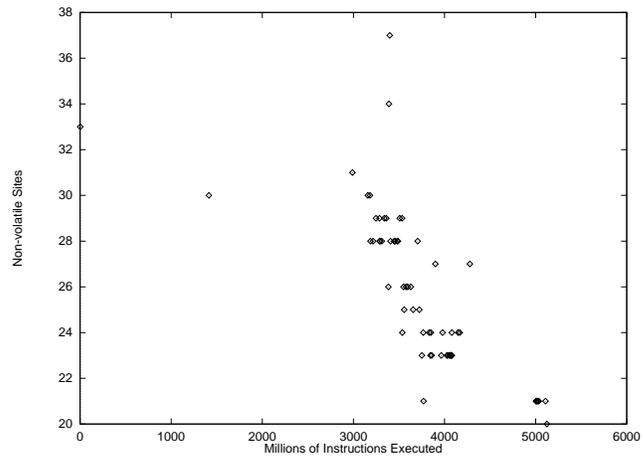}
\end{center}
\caption{Non-volatile site count (complexity estimate) for the set of
  phenotypic Tierran species, as a function of speciation time.}
\label{results}
\end{figure}

Due to the time constraints of producing this paper, the analysis of a
reasonable length Tierra run has not been completed.  At the time of
writing, a moderately large data set of 1660 organisms was generated
from a 24 hour Tierra run. Tierra produces most of its diversity
during the earliest stage of its running, so it becomes significantly
more expensive to produce larger data sets. This data set was halved
by removing every second organism, and then a phenotypic analysis was
carried out. This set reduced to 103 distinct phenotypes, which formed
the test list used for carrying out the complexity analysis. Each of
these 103 organisms were then tested for phenotypic equivalence
against their single site nearest neighbours. The number of sites on
which no mutation resulted in a phenotypically equivalent organism
(``nonvolatile sites'') is plotted against the time of speciation
in figure \ref{results}.

\begin{thebibliography}{1}

\bibitem{McShea96}
McShea, D.W.:
\newblock Metazoan complexity and evolution: Is there a trend?
\newblock Evolution \textbf{50} (1996) 477--492

\bibitem{Adami98}
Adami, C.:
\newblock Introduction to Artificial Life.
\newblock Springer, New York (1998)

\bibitem{Standish97b}
Standish, R.K.:
\newblock Embryology in {T}ierra: A study of a genotype to phenotype map.
\newblock Complexity International \textbf{4} (1997) http://www.csu.edu.au/ci.

\bibitem{Layzer88}
Layzer, D.:
\newblock Growth and order in the universe.
\newblock In Weber, B., Depew, D., Smith, J., eds.: Entropy, Information and
  Evolution.
\newblock MIT Press, Cambridge, Mass. (1988)  23--39

\bibitem{Kauffman95}
Kauffman, S.:
\newblock At Home in the Universe: The Search for Laws of Complexity.
\newblock Oxford UP, New York (1995)

\bibitem{Reidys-etal96}
Reidys, C., Kopp, S., Schuster, P.:
\newblock Evolutionary optimization of biopolymers and sequence structure maps.
\newblock In Langton, C., Shimohara, K., eds.: Artificial Life V, MIT Press
  (1997) 379

\bibitem{Schuster97}
Schuster, P.:
\newblock Landscapes and molecular evolution.
\newblock Physica D \textbf{107} (1997) 351--365

\end{thebibliography}

\end{document}